\documentclass[a4paper]{article}
\usepackage{fullpage}
\usepackage{graphicx}
\usepackage{hyperref} 
\usepackage{mdwlist} 

\begin{document}
\begin{center}

{\large \bf New Variable Stars on Digitized Moscow Collection Plates.\\
Field 66~Ophiuchi (Northern Half)}

\bigskip

\renewcommand{\thefootnote}{\fnsymbol{footnote}}
D.M.~Kolesnikova$^1$, L.A.~Sat$^2$, K.V.~Sokolovsky$^{2,3}$\footnote{Currently at the Max Planck Institute for Radio Astronomy, Auf dem
H\"ugel 69, 53121 Bonn, Germany},
S.V.~Antipin$^{2,1}$, N.N.~Samus$^{1,2}$

\medskip

$^1$Institute of Astronomy, Russian Academy of Sciences, 48,
Pyatnitskaya Str., Moscow 119017, Russia
e-mail: (dkolesnik, samus)@inasan.ru\\
$^2$Sternberg Astronomical Institute, Moscow University, 13,
University Ave., Moscow 119992, Russia
e-mail: lsat@inasan.ru, ksokolov@mpifr-bonn.mpg.de, antipin@sai.msu.ru\\
$^3$Astro Space Center of Lebedev Physical Institute,
Profsoyuznaya 84/32, 117997 Moscow, Russia\\
\bigskip
ABSTRACT
\end{center}

\setcounter{footnote}{0}

We initiated digitization of the Moscow collection of astronomical
plates using flatbed scanners. Techniques of photographic
photometry of the digital images were applied, enabling an
effective search for new variable stars. Our search for new
variables among 140000 stars in the $10^\circ\times5^\circ$
northern half of the field centered at 66~Oph, photographed with
the Sternberg Institute's 40-cm astrograph in 1976--1995, gave
274~new discoveries, among them: 2~probable Population~II
Cepheids; 81~eclipsing variables; 5~high-amplitude $\delta$~Sct
stars (HADSs); 82~RR~Lyr stars; 62~red irregular variables and 41~red
semiregular stars; 1~slow irregular variable not red in color.
Light elements were determined for periodic variable stars. We
detected about 30~variability suspects for follow-up CCD
observations, confirmed 11~stars from the New Catalogue of
Suspected Variable Stars, and derived new light elements for
2~stars already contained in the General Catalogue of Variable
Stars.

\medskip

{\bf Key words:} {\it Stars: variables: general}

\bigskip

\begin{center}

{\bf 1. \ \ Introduction}

\end{center}

Regular photographic observations of the sky for variable-star
studies were started in Moscow in 1895. Since then, several
different telescopes were used to take direct sky plates for
astrometry and for astrophysics. The Moscow plate archive now
contains more than 60000 direct photographs and objective-prism
plates taken in Moscow, at other sites in Russia, and at the
Sternberg Institute's observatory in Crimea, Ukraine.

The most important part of the Moscow plate collection are direct
sky photographs acquired in 1948--1996 with a 40-cm astrograph.
This instrument was ordered by Prof. C.~Hoffmeister for Sonneberg
Observatory (Germany) and first installed there in 1938. 1658
plates from this telescope, taken in 1938--1945, are kept in
Sonneberg (the GA series of the Sonneberg plate collection). In
1945, the telescope was taken to the Soviet Union as a part of the
World War~II reparations. It was initially installed in Simeiz
(Crimea), then brought to Kuchino near Moscow, and in 1958 became
the first instrument of the Crimean Laboratory of the Sternberg
Institute in Nauchny, Crimea. The total number of plates taken
with the 40-cm astrograph after 1948 is about 22500. A single
attempt of direct comparison between Sonneberg and Crimean plates
of the 40-cm astrograph at a blink comparator was undertaken in
1980s (Samus 1983).

The field of view of the 40-cm astrograph is
$10^\circ\times10^\circ$, on $30\times30$~cm plates (the focal
length is 1600~mm). The typical exposure time for the
variable-star fields was 45 minutes. The limiting magnitude of
good-quality plates is about 17$.\!\!^{\rm m}$5~($B$). The
instrument was mainly used for variable-star studies, including
search for new variables. For some fields, rich series of plates
exist (up to $\sim500$ plates). For variable stars that can be
found in several fields, sometimes as many as 1000 photographic
plates are available. The list of fields, with numbers of plates
obtained, can be found in
Internet\footnote{\url{http://cataclysm.sai.msu.ru/www/plates/40.dat}}.
Plates are kept
in good conditions, most plates, initially of excellent quality,
are still perfect.

The Moscow plate collection, like other major astronomical plate
collections of the world, has been actively used for scientific
research for decades. It still contains a large amount of
significant information never used by researchers, as indicated by
discoveries of interesting events missed at the time of
observation, like the discovery of Nova Aql~1985 (V1680~Aql) made
17~years later (Antipin et al. 2002).

Guaranteed conservation of the vast amounts of information
contained in the plate collection and its use by means of modern
methods of image processing require digitization of plate
archives. In Moscow, this work commenced in 2004, after the
purchase of two Creo EverSmart Supreme II scanners. The initial
digitization plans, along with a more detailed description of the
Moscow plate archive from different instruments, were presented in
Samus et al. (2006).

Most plates from the 40-cm astrograph were taken for
variable-star studies. It was natural to
search for new variable stars using digital images obtained in the
process of scanning the Moscow collection plates. In our first
experiments, we discovered 38~new variable objects (mostly
variable stars, but also extragalactic objects) on test partial
scans (several square degrees) of star fields photographed with
the astrograph (Sokolovsky 2006; Manannikov et al. 2006;
Kolesnikova et al. 2007a,b). We introduced preliminary designations
for variable stars discovered in this program with the prefix MDV
(Moscow Digital Variable).

There were several other attempts to search for variable objects
on digitized photographic plates. Among them are: a search for
QSOs on the base of optical variability and zero proper motion
criteria (Scholz et al. 1997;  Brunzendorf and Meusinger 2001), a
search for long-term variability using Sonneberg archival patrol
plates (Vogt et al. 2004), a search for novae in M~31 using
Tautenburg Schmidt plates (Henze et al. 2008).

In this paper, we announce the discovery and study of 274 new MDVs
in the northern half of the field 66~Oph of the 40-cm astrograph.

\medskip

\begin{center}

{\bf 2. \ \ Scanning and Reductions}

\end{center}

\medskip

The field 66~Oph (18$^{\rm h}$00$.\!\!^{\rm m}$3, +$4^\circ22'$,
J2000.0) was photographed with the 40-cm astrograph in 1976--1995,
a total of 254~plates were acquired.

All these plates were scanned with a resolution of 2540~dpi
(1.2~arcseconds per pixel), providing 14 bit per pixel per color.
Color images produced by the scanner were saved in the TIFF (RGB)
format using the scanner software operating in the Mac~OS~X
environment. In our further reductions, we made use only of the
green channel of each image, selected empirically. The files were
then moved to a Linux server equipped with a 5~TB RAID array for
storage and subsequent analysis. The images were converted to the
FITS format using custom-written
software\footnote{\url{ftp://scan.sai.msu.ru/pub/software/tiff2fits}}.
In this paper, we present our analysis of the northern half of the
field ($10^\circ\times5^\circ$) containing about 140000 stars
within our detection limits (see below).

The response to a point source of a given brightness on a
large-scale photographic plate is subject to strong spatial
variations. Obvious reasons for that include aberrations in the
optics of the astrograph (coma, vignetting, etc.), inhomogeneity
in photographic emulsion coating, and differences in airmass for
stars in different parts of a plate. All these factors are
expected to be relatively weak functions of coordinates on a
plate. To overcome these complexities, the $10^\circ\times5^\circ$
field was subdivided into 72~nearly-square subfields. The
influence of systematic factors is assumed to be the same for all
stars in a given subfield. Each subfield was analyzed separately
using VaST\footnote{\url{http://saistud.sai.msu.ru/vast}} software
(Sokolovsky and Lebedev 2005), the results were combined at the
final stage.

For star detection and aperture photometry, VaST uses the
well-known SExtractor code (Bertin and Arnouts 1996). All objects
identified by SExtractor as blended or non-point sources were
excluded from further consideration because such sources produce
many false detections in a variability search. Aperture photometry
was performed with a circular aperture. The aperture diameter was
automatically selected for each image to compensate for seeing
variations. This method was preferred against the variable
elliptical aperture photometry (parameter MAG\_AUTO) enabled by
default in SExtractor, because the addition of extra degrees of
freedom (the aperture shape and size determined for each star
separately) deteriorate the quality of measurements of faint
stars. The SExtractor parameters and the aperture diameter were
selected to optimize measurements of stars in the 13.5--16.5 mag
($B$) range. This magnitude range was
preferred because brighter variable stars in this particular field
have mostly been already discovered in the ASAS-3 (Pojmanski
2002) and ROTSE-I/NSVS (Wo\'zniak et al. 2004) CCD surveys, both
covering the near-equatorial field of our plates.

The VaST code automatically matches stars detected on an image by
SExtractor with stars detected on the reference image using the
technique of the search for similar triangles. One of the best
photographs was chosen as a reference image. Magnitudes of stars
were measured by SExtractor in an instrumental scale with respect
of the background level of the current image. All measured
magnitudes were converted to the instrumental system of the
reference image by approximating the relation between magnitudes
on the current and reference images with a parabolic function. All
stars matched on the images were used to establish this relation.
Visual inspection confirms that this approximation works well in
the required range of magnitudes.

The resulting light curves are characterized by an {\it rms} error of
0.05--0.15 mag for stars in the 13.5--16.5 mag range.

\medskip

\begin{center}

{\bf 3. \ \ The Method of the Search for Variability and Its
Limitations}

\end{center}

\medskip

A light curve of a variable star is, obviously, characterized by a
larger scatter of magnitude measurements compared to non-variable
stars measured on the same series of images. However, the
precision of magnitude measurements for a particular star is a
function not only of its brightness but of many different factors,
like the presence of close companions and image defects. That is
why a variability search based solely on magnitude scatter as a
function of a star's magnitude is inefficient, at least for noisy
photographic data, and will result either in dramatic
incompleteness or in a very large number of false ``positive''
detections. To deal with the problem, we extensively use time
information contained in our data, as described below.

The search for variability in a sample of light curves was
conducted in several steps. First, the relation {\it rms} deviation
-- instrumental magnitude was constructed for each subfield. Stars
with {\it rms} deviations in excess of the average for their
magnitudes were selected using a soft criterion. The second step
was to study time series for each selected star for periodicities
using a number of complementary algorithms:
\begin{itemize*}
\item Our own version of the Phase Dispersion Minimization
algorithm, developed by one of the authors (D.M.K.).
\item An Analysis of Variance (ANOVA, Schwarzenberg-Czerny 1989, 1996)
technique. We made use of the C code from DeBiL package (Devor
2005) implementing this algorithm.
\item Box Least Squares
algorithm (Kov\'acs et al. 2002) originally developed for search
for transiting extrasolar planets. This algorithm has proven to be
useful in identifying Algol-type variables among photographic
light curves.
\end{itemize*}
The listed algorithms provide means to judge on the statistical
significance of detected periodicities. The period significance
cut-offs for candidate selection were chosen for each algorithm
using a number of previously found variable stars.

Along with the periodicity search approach, we used the
variability detection algorithm proposed by Welch and Stetson
(1993) to search for slow (compared to typical time sampling of
our photographic light curves) non-periodic brightness variations
which are often found for post-AGB and AGB stars and for active
galactic nuclei. This technique was used mostly as a complementary
one but not as a main candidate-selection method. Surprisingly, we
found that slow irregular variables could often be detected by
spurious periodicities found by period-search techniques even if
the light variations are non-periodic. These false periods are
usually found around integer multiples of 1~day and they correspond
to beat frequencies between the typical light-curve sampling
frequency and the characteristic frequency of real light
variations. In such cases, visual inspection of a light curve
readily reveals the true character of variability.

Fig.~1 shows the results of our variable-star search in a small
subfield that gave 8~detections of variable stars (some of them
known).

Magnitudes of all detected variable stars were then converted to
the  $B$ scale using a number of USNO-A2.0 stars (Monet et al.
1998). The relation between the instrumental magnitudes and the
USNO-A2.0 $B$ magnitudes for each subfield was, again,
approximated by a parabolic function. This step was performed
after the selection of variable-star candidates since possible
errors on this stage could introduce additional noise into light
curves. A sample calibration diagram for a subfield is displayed
in Fig.~2.

\begin{figure}
\center{\includegraphics[angle=270,width=0.88\textwidth]{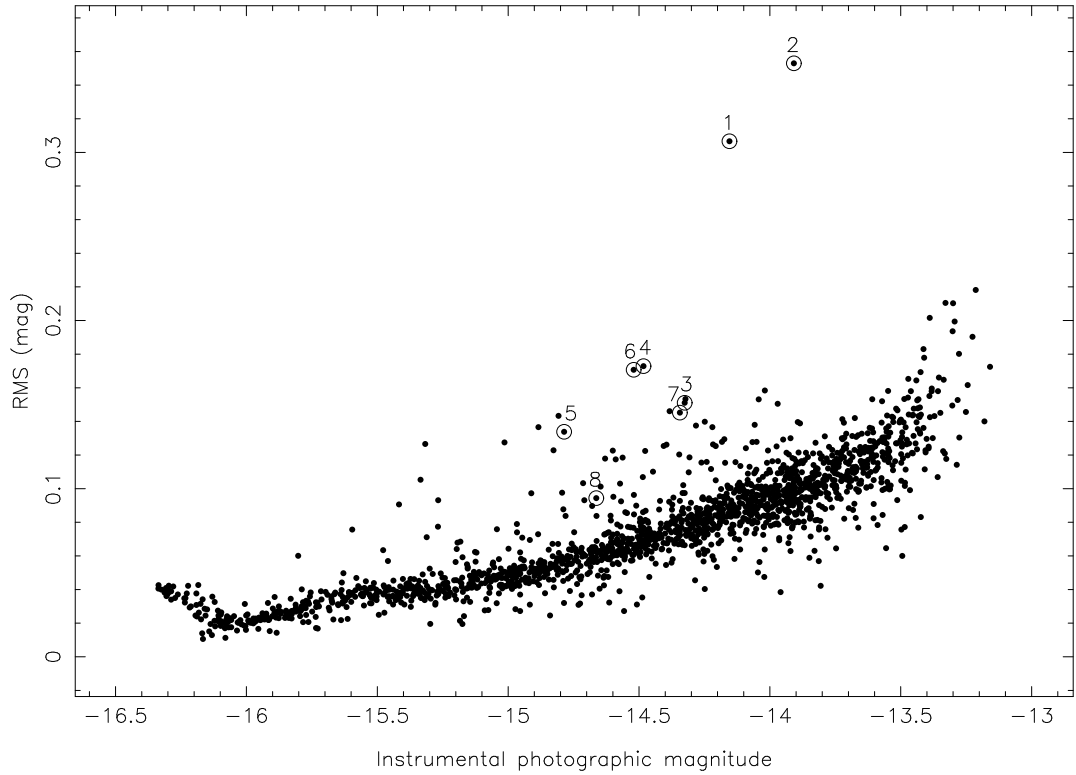}}
\caption{The results of the search for variable stars
in one of the 72~subfields in the northern half of the 66~Oph field.
Circled are the eight detected objects: No.~1 is V1077~Oph, No.~2,
V2328~Oph, No.~3, MDV~92, No.~4, MDV~91, No.~5, MDV~72, No.~6, V940~Oph,
No.~7, MDV~83, No.~8 is one of suspected variables for our future
CCD studies.}
\end{figure}

\begin{figure}
\center{\includegraphics[angle=0,width=0.88\textwidth]{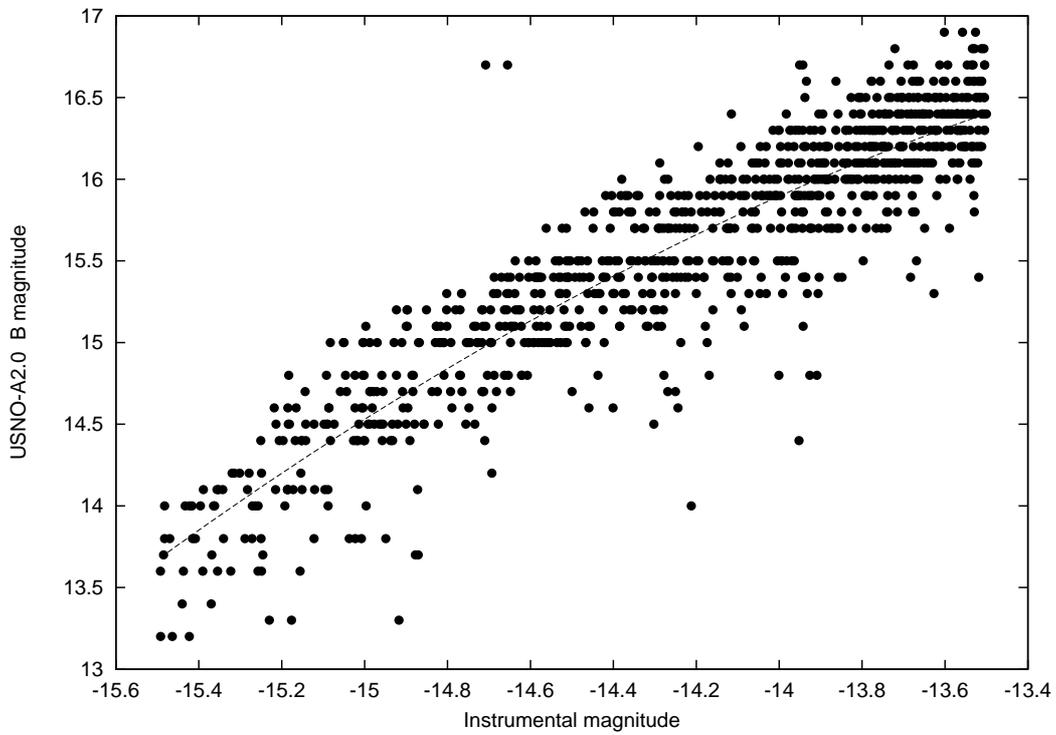}}
\caption{A sample calibration curve for one of subfields. The dashed
curve is the adopted magnitude calibration.}
\end{figure}

Having selected the candidates, we then studied their brightness
variations using the WinEFK software written by Dr. V.P.~Goranskij
and kindly made available to us. This software permits to view
light curves, to look for periodicities using several well-known
algorithms (Deeming, Lafler--Kinman, etc.), to search for second
periodicities. Our final decision if an automatically selected
candidate was a real variable star was made only after a visual
inspection of its light curve.

The described variability search technique has a number of
limitations. First, it is not particularly sensitive to irregular
light variations on time scales shorter than the light curve
sampling time. Objects showing this type of variability can be
detected solely on the base of large magnitude {\it rms} deviations
if a careful inspection of images does not reveal any reason why
this particular star was measured with a much worse precision than
other stars. Without an aid of the period search technique, this
results in a much worse detection probability for such variations.
No such objects were found in the field described in this paper.
However, T~Tau variables found during a special search in the
field of V451~Tau show exactly this behavior. The results of the
variability search in the V451~Tau field will be discussed
elsewhere.

The second limitation results from the properties of the VaST
software. This software constructs light curves only for those
stars detected on the reference image for which the total number
of detections exceeds 30. This approach effectively avoids many
false star detections (because of plate flaws, dust, and large
grains of the emulsion) but remains sensitive even to the faintest
stars visible on the plates. However, this makes us completely
insensitive to any transient phenomena (Novae, dwarf nova
outbursts, etc.) that can be present on the plates.

\medskip

\begin{center}

{\bf 4. \ \ Results}

\end{center}

\medskip

As expected, we detected rather many known variable stars. They
were analyzed along with the new variables (see below), but this
paper deals with only those of them for which our results
significantly correct or append published information.

We have discovered a total of 274~new variable stars
(MDV~39 -- MDV~312). They are presented in Table~1. Among these
stars, there are 2~probable Population~II Cepheids; 81~eclipsing
variables; 5~high-amplitude $\delta$~Sct stars (HADSs);
82~RR~Lyr stars; 62~red irregular variables and 41~red semiregular
stars; 1~slow irregular variable not red in color (MDV~80).

\begin{figure}
\center{\includegraphics[angle=0,width=1.0\textwidth]{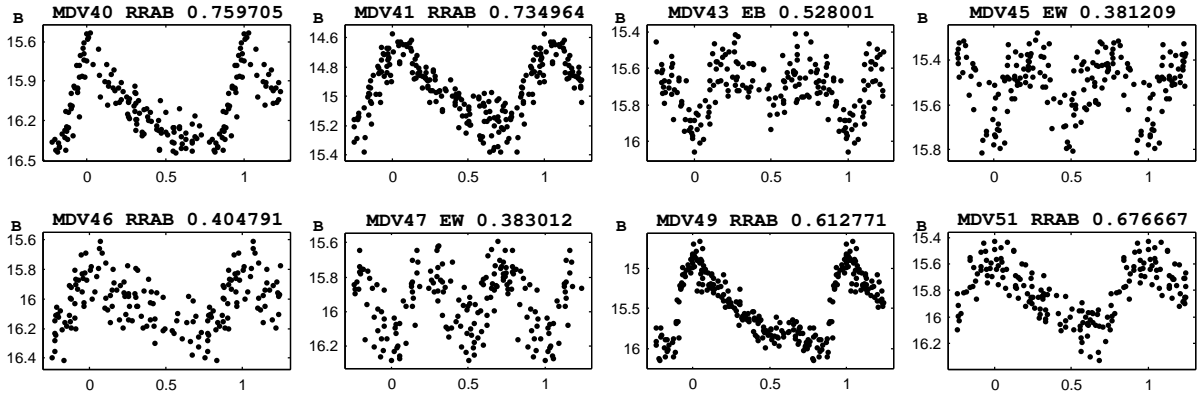}}
\caption{Sample phased light curves of the new regular
variable stars. Only the first 8 light curves are shown.}
\end{figure}

Our phased photographic light curves of the new periodic variable
stars (with the exception of some of the red semiregular
variables) can be found at the web site of our
team\footnote{\url{http://www.sai.msu.su/gcvs/digit/mdv/}}. Fig.~3 shows, as an
example, the first eight phased light curves. Fig.~4 is the
light curve of MDV~80. The observations of all the new variable
stars are also available at our web
site\footnote{\url{http://www.sai.msu.su/gcvs/digit/mdv/data/}}.

\begin{figure}
\center{\includegraphics[angle=0,width=1.0\textwidth]{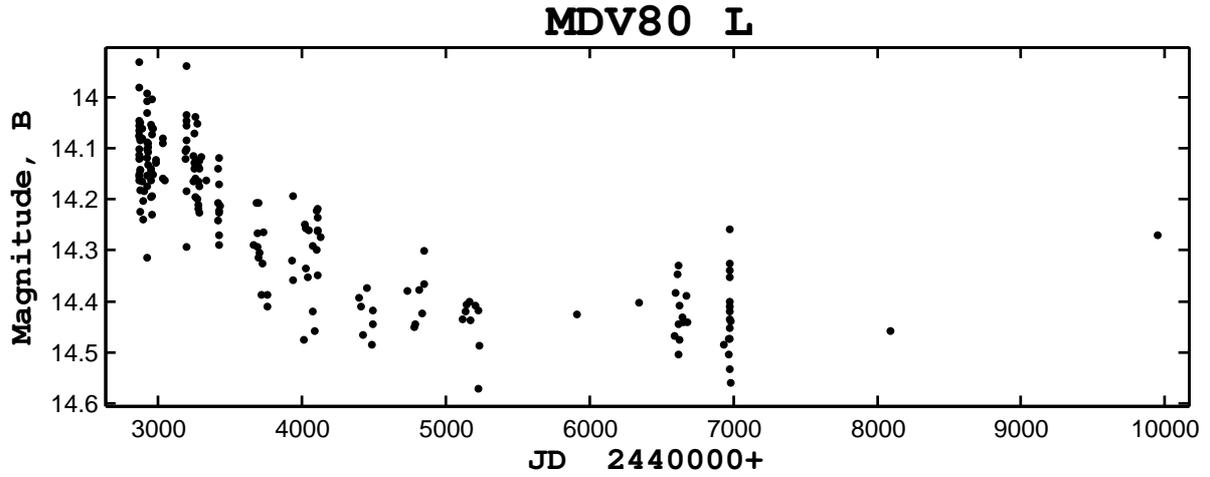}}
\caption{The light curve of the ``white'' slow irregular
variable star MDV~80.}
\end{figure}

\begin{figure}
\center{\includegraphics[angle=0,width=1.0\textwidth]{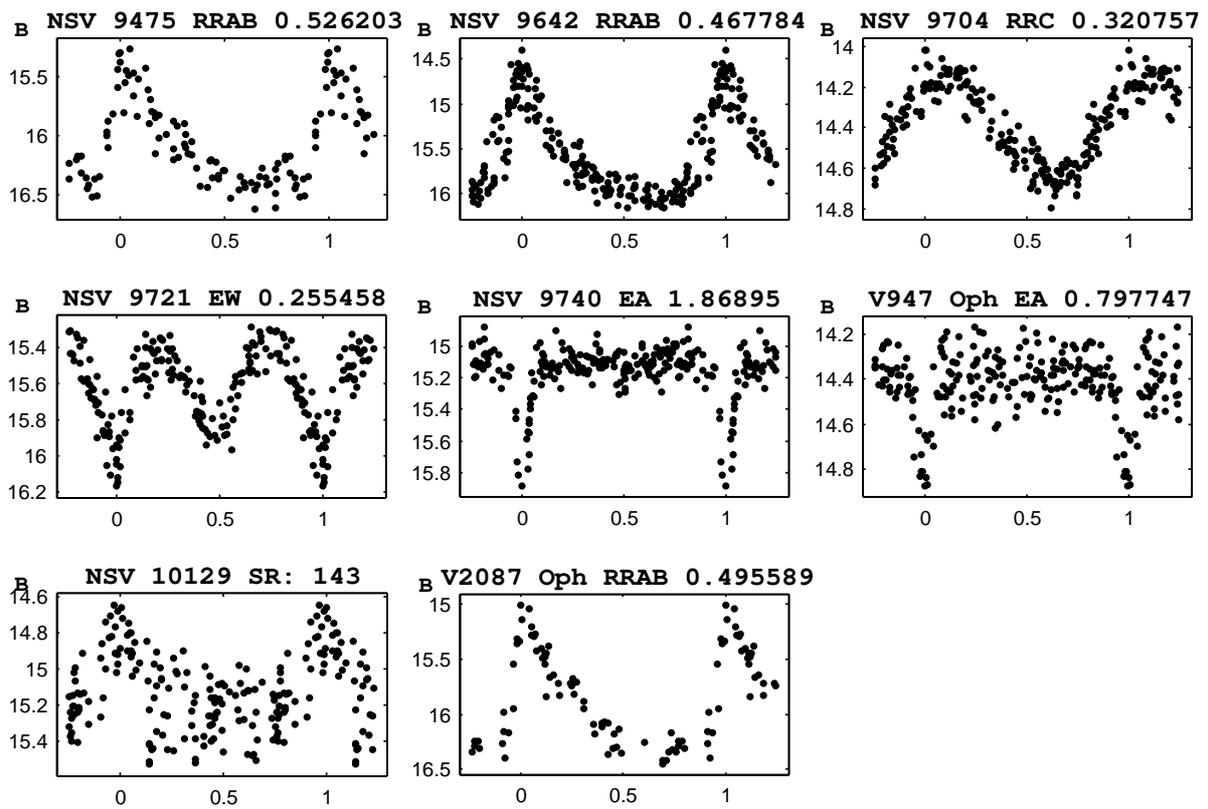}}
\caption{Phased light curves for the known regular
variable stars in the field that were investigated in this study.}
\end{figure}

Most new red variable stars we have detected also definitely vary
in the NSVS observations. It should be specially noted that we did
not use the NSVS data to discover new variables in our field but
attracted them only for independent confirmation of our
discoveries.

Table~1 does not include some 30~stars we have selected as
variability suspects. Their amplitudes are too small for reliable
judgment from photographic data if they are genuine variable
stars. We are planning special follow-up CCD observations to
confirm and study these variables.

We were able to confirm variability of 11~stars earlier in the
NSV~catalogue. Six of them are periodic variables, for them we
present the light elements for the first time. The remaining
five stars are red irregular variables. We also determined new light
elements for two GCVS stars: the period we find for the eclipsing
star V947~Oph is completely different from that in G\"otz et al.
(1957), and we present the first light elements for the RR~Lyr
variable V2087~Oph. The data on the 13~known variables are
presented in Table~2. The coordinates given in Tables~1 and~2,
unless stated, are from the 2MASS point-source catalogue (Cutri et
al. 2003). The light curves for the eight periodic stars are
displayed in Fig.~5.

\medskip

\begin{center}

{\bf 5. \ \ Conclusions}

\end{center}

\medskip

We have successfully developed necessary techniques to digitize
plates of the Moscow collection, search for variable stars using
digital images, perform photographic photometry. These techniques
will be further improved and repeatedly used in our future
research. Its results will be published separately and
presented at our web
site\footnote{\url{http://www.sai.msu.su/gcvs/digit/digit.html}}.

This study resulted in the discovery and investigation of 274~new
variable stars of different types, periodic and aperiodic, fast
and slow, in a $10^\circ\times5^\circ$ field, demonstrating the
effectiveness of our approach. Additionally, we found about
30~variability suspects for follow-up CCD studies, confirmed
variability of 11~stars from the NSV catalog, and determined light
elements for 2~GCVS stars. All these results were achieved for a
field rather well-studied for stellar variability before.

\medskip

{\bf Acknowledgments.}
We wish to thank D.~Nasonov, S.~Nazarov, and especially A.~Lebedev
for their contribution to development of VaST software. Thanks are
due to A.~Belinsky for his support of our scanning project. We are
grateful to V.P.~Goranskij for providing us access to his software
for periodicity analysis. This research has made use of the SIMBAD
database, operated at CDS, Strasbourg, France and the
International Variable Star Index (VSX) operated by the AAVSO. Our
study was supported, in part, by the Russian Foundation for Basic
Research (grants 05-02-16688 and 08-02-00375) and by the Program
of Support for Leading Scientific Schools of Russia (grant
NSh-433.2008.2).

\newpage

\hoffset=-0.06\textwidth

\begin{center}

Table 1. New Moscow Digital Variables

\medskip
{\small
\begin{tabular}{|r|c|l|l|l|cl|l|l|}
\hline MDV & Coord. (J2000) & GSC / USNO-A2.0 & type &
max-min-min II &\multicolumn{2}{|c|}{epoch JD24...} & period & rem. \\
\hline
  39 & 17 40 07.83 +4 24 28.0 & A2 0900-10266740 & LB     & 16.0-16.6       &     &           &             & 1       \\
  40 & 17 40 29.79 +6 55 42.1 & A2 0900-10281306 & RRAB   & 15.6-16.35      & max & 44847.280 &   0.759705  &         \\
  41 & 17 40 35.50 +6 17 00.4 & A2 0900-10285157 & RRAB   & 14.65-15.3      & max & 42902.512 &   0.734964  &         \\
  42 & 17 41 18.87 +3 49 25.6 & GSC 00423-01670  & SR:    & 15.1-15.6       &     &           &             & 1, 2    \\
  43 & 17 42 13.95 +5 42 45.2 & A2 0900-10351850 & EB     & 15.6-16.0-15.85 & min & 44491.256 &   0.528001  & 3       \\
  44 & 17 42 21.25 +4 18 42.2 & GSC 00423-00845  & SRB    & 14.8-15.15      &     &           &             & 1, 4    \\
  45 & 17 42 44.86 +4 46 37.4 & A2 0900-10372558 & EW     & 15.4-15.75-15.7 & min & 43282.452 &   0.381209  &         \\
  46 & 17 43 04.48 +6 54 44.4 & A2 0900-10385352 & RRAB   & 15.7-16.35      & max & 43285.493 &   0.404791  &         \\
  47 & 17 43 52.96 +6 36 02.3 & A2 0900-10416750 & EW     & 15.7-16.2-16.2  & min & 42957.469 &   0.383012  &         \\
  48 & 17 45 13.67 +5 03 15.6 & GSC 00424-00974  & LB     & 15.7-16.8       &     &           &             & 1       \\
  49 & 17 45 18.14 +8 14 20.9 & A2 0975-09664523 & RRAB   & 14.8-16.0       & max & 42922.490 &   0.612771  &         \\
  50 & 17 45 49.57 +5 13 49.9 & GSC 00424-00684  & LB     & 14.05-14.45     &     &           &             & 5       \\
  51 & 17 45 57.40 +5 02 45.6 & A2 0900-10497630 & RRAB   & 15.6-16.2       & max & 44072.391 &   0.676667  &         \\
  52 & 17 46 15.35 +6 14 14.5 & A2 0900-10509345 & SRB    & 15.4-16.1       &     &           &  58.5:      & 1, 6    \\
  53 & 17 47 23.26 +6 10 26.7 & GSC 00428-00825  & EW     & 15.2-15.8-15.7  & min & 42876.527 &   0.353475  &         \\
  54 & 17 47 55.07 +8 15 58.3 & A2 0975-09739370 & EW     & 14.9-15.4-15.35 & min & 44027.451 &   0.315782  &         \\
  55 & 17 48 09.07 +5 08 33.3 & A2 0900-10583670 & RRAB   & 16.1-17.0       & max & 43283.447 &   0.641829  &         \\
  56 & 17 48 09.74 +6 09 09.8 & A2 0900-10584141 & EW     & 15.75-16.3-16.25& min & 44025.432 &   0.254585  &         \\
  57 & 17 48 23.11 +6 37 29.5 & GSC 00428-01925  & SR:    & 14.9-15.25      &     &           &  48:        & 1, 7    \\
  58 & 17 48 29.26 +5 29 16.8 & A2 0900-10597576 & EA     & 16.0-16.5       & min & 46344.23  &   0.98930   & 8       \\
  59 & 17 48 34.33 +6 30 25.4 & GSC 00428-00148  & EB     & 14.55-15.0-14.8 & min & 45941.312 &   0.473063  &         \\
  60 & 17 48 34.57 +8 09 40.9 & A2 0975-09759232 & RRC    & 15.8-16.2       & max & 42933.452 &   0.310296  &         \\
  61 & 17 48 40.02 +8 29 06.0 & A2 0975-09761937 & RRAB   & 14.8-16.0       & max & 42892.539 &   0.640799  &         \\
  62 & 17 48 50.05 +5 59 20.3 & A2 0900-10611848 & LB     & 15.4-16.0       &     &           &             & 1       \\
  63 & 17 49 11.11 +8 24 26.2 & A2 0975-09777699 & RRC:   & 14.7-15.2       & max & 43253.517 &   0.443354  & 9       \\
  64 & 17 49 11.11 +6 03 53.0 & A2 0900-10627092 & RRAB   & 15.6-16.4       & max & 42933.452 &   0.614422  &         \\
  65 & 17 49 37.25 +6 51 44.2 & GSC 00428-00414  & SRB:   & 15.15-15.6      &     &           &  54.9:      & 1       \\
  66 & 17 49 46.77 +5 30 56.5 & A2 0900-10653069 & RRAB   & 14.15-14.5      & max & 42872.494 &   0.721473  &         \\
  67 & 17 49 51.67 +7 16 41.1 & A2 0900-10656595 & EW     & 15.9-16.2-16.2  & min & 42876.562 &   0.378639  &         \\
  68 & 17 50 46.76 +6 07 57.8 & A2 0900-10697991 & RRC    & 15.4-16.0       & max & 43702.392 &   0.392512  &         \\
  69 & 17 50 50.76 +7 51 10.8 & A2 0975-09829154 & SRB    & 15.8-16.2       &     &           &  47:        & 1, 10   \\
  70 & 17 50 56.82 +6 21 19.7 & GSC 00428-01901  & LB     & 15.0-15.7       &     &           &             & 5       \\
  71 & 17 51 23.46 +8 25 02.5 & GSC 00994-01460  & RRAB:  & 15.25-15.7      & max & 42954.322 &   0.392501  &         \\
  72 & 17 51 34.63 +7 40 30.7 & A2 0975-09851918 & RRC    & 14.95-15.5      & max & 44043.431 &   0.322902  &         \\
  73 & 17 51 37.90 +8 44 01.5 & A2 0975-09853705 & HADS   & 15.3-15.9       & max & 43198.598 &   0.0999541 & 11      \\
  74 & 17 51 45.02 +3 47 54.8 & A2 0900-10744203 & LB     & 15.5-16.1       &     &           &             &         \\
  75 & 17 51 48.78 +9 26 33.6 & A2 0975-09859679 & RRAB   & 14.2-15.6       & max & 43700.317 &   0.487961  &         \\
  76 & 17 51 54.83 +5 26 14.8 & GSC 00424-00123  & LB     & 15.1-16.1       &     &           &             & 1       \\
  77 & 17 51 59.69 +3 56 22.6 & A2 0900-10756126 & RRC    & 15.1-15.7       & max & 44087.407 &   0.318240  &         \\
  78 & 17 52 02.92 +8 49 51.6 & A2 0975-09867207 & EW     & 15.0-15.6-15.5  & min & 43190.597 &   0.412666  &         \\
  79 & 17 52 05.13 +4 33 22.1 & A2 0900-10760510 & EA     & 15.0-15.6       & min & 49949.335 &   1.68149   & 8       \\
  80 & 17 52 24.09 +4 31 57.8 & GSC 00424-01416  & L      & 14.0-14.5       &     &           &             & 12      \\
  81 & 17 52 24.75 +9 16 16.1 & A2 0975-09878755 & RRC    & 15.4-15.85      & max & 44087.407 &   0.292674  & 9       \\
  82 & 17 52 31.50 +5 01 18.0 & A2 0900-10781877 & RRC    & 16.0-16.4       & max & 42871.515 &   0.336160  &         \\
  83 & 17 52 37.26 +7 20 49.9 & A2 0900-10786611 & EW     & 15.45-15.9-15.9 & min & 42894.525 &   0.355982  &         \\
  84 & 17 52 40.57 +8 28 06.2 & A2 0975-09887194 & EW     & 15.7-16.05-16.0 & min & 44077.360 &   0.429948  &         \\
  85 & 17 52 53.84 +4 24 34.4 & A2 0900-10800230 & RRC    & 15.9-16.2       & max & 42875.563 &   0.256808  &         \\
  86 & 17 53 04.03 +6 14 32.0 & A2 0900-10808592 & SR:    & 15.8-16.45      &     &           & 143:        & 1       \\
  87 & 17 53 04.77 +6 13 45.7 & A2 0900-10763277 & RRAB   & 16.1-16.6       & max & 42876.562 &   0.584480  &         \\
  88 & 17 53 08.46 +4 32 04.2 & A2 0900-10812302 & RRC    & 15.7-16.2       & max & 43289.393 &   0.284575  &         \\
  89 & 17 53 22.29 +4 00 01.1 & A2 0900-10823986 & EW     & 14.4-14.95-14.9 & min & 44012.480 &   0.677689  &         \\
  90 & 17 53 32.34 +6 08 34.7 & GSC 00429-02191  & EW     & 14.2-14.65-14.6 & min & 46973.322 &   0.299660  &         \\
  91 & 17 53 44.10 +7 00 52.6 & GSC 00429-02060  & LB     & 15.2-16.0       &     &           &             & 1       \\
  92 & 17 53 50.57 +7 13 27.1 & A2 0900-10847301 & RRC    & 15.5-15.95      & max & 46934.425 &   0.312434  &         \\
  93 & 17 53 52.77 +6 11 25.7 & GSC 00429-01936  & LB:    & 15.0-15.45      &     &           &             & 1, 13   \\
  94 & 17 54 23.42 +8 13 13.0 & A2 0975-09945912 & RRAB   & 15.2-16.1       & max & 44491.256 &   0.492470  &         \\
  95 & 17 54 23.45 +6 21 45.3 & A2 0900-10874520 & RRAB   & 15.4-16.25      & max & 42922.490 &   0.603587  &         \\
\hline
\end{tabular}
}

\newpage

Table 1. Continued

\medskip

{\small
\begin{tabular}{|r|c|l|l|l|cl|l|l|}
\hline MDV & Coord. (J2000) & GSC / USNO-A2.0 & type &
max-min-min II &\multicolumn{2}{|c|}{epoch JD24...} & period & rem. \\
\hline
  96 & 17 54 24.07 +5 05 44.6 & GSC 00425-01277  & CWB:   & 15.1-15.6       & max & 46979.46  &   4.22851   & 14      \\
  97 & 17 54 34.54 +4 31 43.4 & A2 0900-10884039 & RRAB   & 15.7-16.7       & max & 46618.465 &   0.555057  &         \\
  98 & 17 55 12.15 +8 30 49.7 & A2 0975-09978626 & RRC    & 15.9-16.4       & max & 42894.525 &   0.274916  &         \\
  99 & 17 55 14.25 +7 55 00.6 & A2 0975-09980031 & EB     & 14.95-15.25-15.1& min & 43426.226 &   0.580428  &         \\
 100 & 17 55 22.79 +4 00 23.2 & A2 0900-10928663 & RRC    & 14.25-14.8      & max & 46972.316 &   0.312420  &         \\
 101 & 17 55 34.54 +7 19 33.0 & A2 0900-10939727 & EW     & 15.3-15.9-15.85 & min & 44397.415 &   0.545532  & 15      \\
 102 & 17 55 36.01 +9 04 14.8 & A2 0975-09994108 & RRAB   & 15.4-16.3       & max & 43253.517 &   0.761835  &         \\
 103 & 17 55 44.14 +5 59 29.5 & A2 0900-10948278 & RRAB:  & 15.9-16.2       & max & 45941.312 &   0.945095  &         \\
 104 & 17 55 50.95 +8 11 09.9 & A2 0975-10003602 & EW     & 15.75-16.1-16.0 & min & 42963.332 &   0.542425  &         \\
 105 & 17 55 59.58 +4 02 28.8 & A2 0900-10962191 & SR     & 15.7-16.4       &     &           & 145         & 1       \\
 106 & 17 56 05.81 +8 11 35.3 & A2 0975-10012892 & CWA:   & 15.4-15.8       & max & 42894.53  &  16.56      &         \\
 107 & 17 56 17.88 +6 22 43.0 & A2 0900-10977879 & HADS   & 14.65-15.3      & max & 43249.548 &   0.107927  &         \\
 108 & 17 56 41.05 +5 53 58.0 & GSC 00429-00460  & LB     & 14.7-15.3       &     &           &             & 1       \\
 109 & 17 56 50.87 +8 08 02.0 & A2 0975-10041649 & RRC    & 15.7-16.05      & max & 44397.415 &   0.350064  &         \\
 110 & 17 56 56.53 +4 19 08.5 & GSC 00425-00661  & LB     & 14.6-15.1       &     &           &             & 1       \\
 111 & 17 57 01.57 +6 15 24.6 & A2 0900-11015717 & LB     & 15.85-16.3      &     &           &             & 1       \\
 112 & 17 57 32.53 +5 19 20.2 & A2 0900-11043387 & RRC:   & 15.5-16.2       & max & 43427.283 &   0.332701  &         \\
 113 & 17 57 42.66 +4 23 59.1 & A2 0900-11052443 & RRAB   & 15.8-16.6       & max & 44789.394 &   0.464905  &         \\
 114 & 17 57 44.84 +7 32 52.6 & A2 0975-10077478 & RRC:   & 16.15-16.4      & max & 42963.332 &   0.284003  &         \\
 115 & 17 57 46.82 +5 55 29.1 & A2 0900-11056158 & EW     & 14.9-15.6-15.5  & min & 46972.320 &   0.547264  & 3       \\
 116 & 17 57 58.89 +4 09 47.6 & A2 0900-11066904 & EB     & 15.9-16.5-16.1  & min & 43034.230 &   0.913185  &         \\
 117 & 17 58 01.29 +5 49 26.7 & GSC 00429-01622  & EB     & 13.65-14.4-13.8 & min & 42957.338 &   0.425898  &         \\
 118 & 17 58 05.41 +9 02 33.3 & A2 0975-10091473 & RRAB   & 15.1-16.2       & max & 43938.578 &   0.777133  &         \\
 119 & 17 58 07.94 +8 22 59.0 & A2 0975-10093169 & RRAB   & 15.3-16.0       & max & 43189.593 &   0.633042  &         \\
 120 & 17 58 09.44 +8 05 09.5 & A2 0975-10094181 & RRAB   & 15.6-16.5       & max & 44455.302 &   0.477403  &         \\
 121 & 17 58 21.36 +6 35 08.4 & A2 0900-11087520 & LB     & 15.9-16.5       &     &           &             & 5       \\
 122 & 17 58 22.55 +5 53 15.2 & GSC 00429-00150  & EB     & 14.4-14.9-14.65 & min & 44494.247 &   0.323388  &         \\
 123 & 17 58 28.58 +6 10 01.1 & A2 0900-11094464 & LB     & 15.7-16.3       &     &           &             & 1       \\
 124 & 17 58 34.10 +6 02 56.6 & GSC 00429-00842  & LB:    & 15.0-15.9       &     &           &             & 1, 16   \\
 125 & 17 58 34.73 +5 01 06.0 & GSC 00425-00007  & LB     & 15.4-16.5       &     &           &             & 1       \\
 126 & 17 58 40.05 +4 54 16.7 & A2 0900-11105212 & EB     & 15.45-15.9-15.7 & min & 42871.520 &   0.649782  &         \\
 127 & 17 58 48.46 +7 16 42.6 & A2 0900-11113082 & EA     & 15.3-15.9       & min & 42872.52  &   1.55847   &         \\
 128 & 17 58 59.30 +4 20 03.2 & A2 0900-11123180 & RRC    & 15.6-16.2       & max & 42922.490 &   0.318133  & 9       \\
 129 & 17 59 11.58 +9 01 53.0 & A2 0975-10136593 & EA     & 15.0-16.0       & min & 43272.41  &   6.3593    &         \\
 130 & 17 59 12.18 +7 52 05.4 & GSC 01007-01100  & LB     & 14.3-14.7       &     &           &             & 17      \\
 131 & 17 59 15.95 +5 13 13.7 & A2 0900-11138974 & RRC    & 16.0-16.5       & max & 46977.463 &   0.386846  &         \\
 132 & 17 59 19.71 +7 51 04.9 & A2 0975-10142159 & RRAB   & 15.0-16.0       & max & 43284.449 &   0.599402  &         \\
 133 & 17 59 22.50 +5 07 33.1 & A2 0900-11145400 & EB     & 15.55-16.0-15.8 & min & 42989.295 &   0.404381  &         \\
 134 & 17 59 29.36 +4 32 33.2 & GSC 00425-01015  & LB     & 15.2-15.65      &     &           &             & 1       \\
 135 & 17 59 29.41 +8 45 41.8 & A2 0975-10148658 & EA     & 14.8-15.4-15.0  & min & 46977.46  &   1.20425   &         \\
 136 & 17 59 39.81 +4 59 51.3 & GSC 00425-00040  & SRB    & 13.7-14.2       &     &           &  66.5:      & 1, 18   \\
 137 & 17 59 47.76 +9 22 41.9 & A2 0975-10161161 & RRC    & 15.3-15.9       & max & 43289.393 &   0.285936  & 9       \\
 138 & 17 59 48.51 +8 10 48.3 & GSC 01007-01237  & EW     & 13.7-14.2-14.15 & min & 45203.305 &   0.345281  &         \\
 139 & 18 00 00.75 +7 26 22.0 & GSC 00442-00127  & LB     & 14.6-15.0       &     &           &             & 1       \\
 140 & 18 00 03.71 +4 41 21.4 & A2 0900-11185755 & RRAB   & 15.4-16.3       & max & 42872.523 &   0.684805  &         \\
 141 & 18 00 07.82 +4 30 27.0 & A2 0900-11189631 & SR:    & 14.8-15.3       &     &           &  31:        & 1       \\
 142 & 18 00 12.49 +6 26 26.1 & A2 0900-11194031 & EW     & 15.0-15.4-15.35 & min & 44847.280 &   0.448969  &         \\
 143 & 18 00 32.68 +6 50 24.0 & GSC 00442-01610  & LB     & 15.1-15.5       &     &           &             & 1       \\
 144 & 18 00 37.01 +8 55 07.7 & GSC 01008-00060  & RRAB   & 14.4-14.8       & max & 42874.530 &   0.66754   &         \\
 145 & 18 00 37.83 +5 06 00.7 & A2 0900-11218671 & EA     & 15.5-16.4-15.9  & min & 44847.28  &   2.10038   &         \\
 146 & 18 00 56.85 +9 21 29.9 & A2 0975-10208777 & LB     & 15.5-16.2       &     &           &             & 1       \\
 147 & 18 00 59.38 +7 21 22.4 & A2 0900-11240189 & RRC:   & 16.25-16.5      & max & 45228.243 &   0.330633  &         \\
 148 & 18 01 00.83 +4 07 51.8 & A2 0900-11241662 & EW     & 15.3-16.1-15.9  & min & 44850.280 &   0.324859  &         \\
 149 & 18 01 05.60 +6 21 14.1 & A2 0900-11246429 & RRAB   & 14.4-15.3       & max & 43249.546 &   0.581607  &         \\
 150 & 18 02 03.93 +7 08 06.6 & A2 0900-11304795 & EW     & 15.65-15.9-15.9 & min & 46619.406 &   0.332680  & 19      \\
 151 & 18 02 11.42 +7 26 42.7 & GSC 00442-00055  & LB     & 14.5-15.0       &     &           &             & 1       \\
 152 & 18 02 12.50 +6 48 14.4 & A2 0900-11313535 & SRB:   & 15.2-15.9       &     &           &  92:        & 1, 20   \\
\hline
\end{tabular}
}

\newpage

Table 1. Continued

\medskip

{\small
\begin{tabular}{|r|c|l|l|l|cl|l|l|}
\hline MDV & Coord. (J2000) & GSC / USNO-A2.0 & type &
max-min-min II &\multicolumn{2}{|c|}{epoch JD24...} & period & rem. \\
\hline
 153 & 18 02 13.54 +6 52 59.5 & A2 0900-11314540 & EB     & 15.6-16.1-15.95 & min & 42925.456 &   0.464174  &         \\
 154 & 18 02 14.03 +8 12 18.4 & A2 0975-10263405 & EW     & 16.0-16.5-16.4  & min & 43685.342 &   0.414680  &         \\
 155 & 18 02 23.14 +8 01 39.6 & A2 0975-10270079 & RRAB   & 14.0-14.9       & max & 43036.237 &   0.679221  &         \\
 156 & 18 02 29.47 +7 12 47.6 & A2 0900-11330921 & RRAB   & 14.8-16.2       & max & 42963.505 &   0.509009  &         \\
 157 & 18 02 36.98 +5 32 12.0 & A2 0900-11338540 & RRAB   & 14.6-15.5       & max & 42922.490 &   0.537830  & 21      \\
 158 & 18 02 41.85 +4 33 02.0 & GSC 00438-02006  & EW     & 13.8-14.15-14.15& min & 45171.387 &   0.381728  &         \\
 159 & 18 02 49.98 +8 43 57.7 & A2 0975-10289889 & EW     & 15.1-15.6-15.6  & min & 46596.478 &   0.392278  &         \\
 160 & 18 02 54.90 +6 34 10.1 & A2 0900-11357287 & EW     & 14.05-14.6-14.6 & min & 43277.523 &   0.376706  &         \\
 161 & 18 02 57.07 +6 02 09.7 & A2 0900-11359572 & EB     & 15.8-16.1-15.95 & min & 43692.392 &   0.392170  &         \\
 162 & 18 02 59.80 +7 22 02.3 & GSC 00442-00400  & LB     & 15.0-15.5       &     &           &             & 1       \\
 163 & 18 03 05.98 +5 03 01.3 & GSC 00438-01095  & SRB    & 13.9-14.3       &     &           &  63.5:      & 1, 22   \\
 164 & 18 03 07.17 +4 20 41.4 & A2 0900-11369966 & LB     & 15.4-15.8       &     &           &             & 5       \\
 165 & 18 03 10.52 +3 57 14.6 & A2 0900-11373534 & EW     & 14.2-14.65-14.65& min & 43046.268 &   0.307384  &         \\
 166 & 18 03 24.13 +5 16 56.1 & GSC 00438-01591  & LB     & 15.2-15.8       &     &           &             &         \\
 167 & 18 03 24.67 +6 40 58.6 & GSC 00442-01740  & LB     & 14.7-15.1       &     &           &             & 1       \\
 168 & 18 03 51.48 +8 05 36.6 & A2 0975-10335151 & EB     & 15.6-16.2-16.05 & min & 44732.521 &   0.399152  &         \\
 169 & 18 04 04.70 +3 54 19.7 & GSC 00438-00764  & LB     & 14.6-15.15      &     &           &             & 1       \\
 170 & 18 04 16.53 +8 23 39.1 & GSC 01008-01224  & LB     & 14.6-15.15      &     &           &             & 1       \\
 171 & 18 04 22.43 +8 12 34.0 & GSC 01008-01677  & SRB    & 14.9-15.7       &     &           &  74         & 1, 23   \\
 172 & 18 04 34.12 +4 43 20.6 & A2 0900-11460939 & EW     & 15.15-15.6-15.5 & min & 43198.600 &   0.335051  & 24      \\
 173 & 18 04 39.62 +7 31 03.3 & A2 0975-10371582 & RRAB   & 15.1-16.1       & max & 42891.529 &   0.589471  &         \\
 174 & 18 04 43.43 +5 53 27.3 & A2 0900-11470688 & RRAB   & 15.2-16.1       & max & 42926.500 &   0.658642  &         \\
 175 & 18 04 51.24 +8 03 57.8 & A2 0975-10380595 & RRC    & 15.2-15.8       & max & 43420.247 &   0.276090  &         \\
 176 & 18 04 58.48 +3 46 03.9 & GSC 00438-00473  & EB     & 13.9-14.3-14.15 & min & 43332.356 &   0.749889  &         \\
 177 & 18 05 00.41 +6 22 51.1 & GSC 00442-00933  & LB     & 15.5-16.05      &     &           &             & 1       \\
 178 & 18 05 27.92 +7 16 51.1 & GSC 00442-00860  & LB     & 14.3-15.1       &     &           &             & 1       \\
 179 & 18 05 29.88 +6 07 53.3 & A2 0900-11519382 & EB     & 15.15-15.8-15.3 & min & 43272.375 &   1.018765  &         \\
 180 & 18 05 35.60 +8 35 00.0 & A2 0975-10414913 & RRAB   & 15.0-16.3       & max & 42951.355 &   0.531888  &         \\
 181 & 18 05 39.42 +5 10 11.8 & A2 0900-11529745 & HADS   & 15.7-16.1       & max & 43418.213 &   0.131870  & 21      \\
 182 & 18 05 42.75 +5 20 18.1 & GSC 00438-01029  & EB     & 14.25-15.0-14.5 & min & 42891.529 &   0.502317  & 25      \\
 183 & 18 05 49.45 +7 55 12.8 & A2 0975-10425490 & SRB:   & 15.2-16.1       &     &           &  70:        & 1       \\
 184 & 18 05 54.43 +7 13 48.2 & A2 0900-11545785 & RRAB   & 15.9-16.8       & max & 42874.564 &   0.617842  &         \\
 185 & 18 06 05.81 +5 54 54.4 & A2 0900-11558043 & RRAB   & 14.8-15.2       & max & 42957.403 &   0.657303  &         \\
 186 & 18 06 12.13 +5 06 37.8 & A2 0900-11565052 & RRAB:  & 15.5-16.2       & max & 42868.507 &   0.923155  &         \\
 187 & 18 06 15.28 +7 06 35.1 & A2 0900-11568523 & RRC    & 16.0-16.7       & max & 43417.212 &   0.332192  & 26      \\
 188 & 18 06 18.63 +8 55 20.5 & A2 0975-10449092 & EB     & 14.9-15.4-15.1  & min & 43199.585 &   0.740361  &         \\
 189 & 18 06 30.80 +8 22 06.0 & A2 0975-10458817 & EB     & 15.8-16.25-16.1 & min & 42957.470 &   0.379855  &         \\
 190 & 18 06 31.82 +3 59 52.8 & A2 0900-11586690 & SR:    & 15.5-16.1       &     &           &  47:        & 1       \\
 191 & 18 06 32.78 +4 29 32.2 & A2 0900-11587789 & RRAB   & 14.8-15.8       & max & 42925.392 &   0.670269  &         \\
 192 & 18 06 36.54 +5 00 43.6 & A2 0900-11592041 & RRAB   & 15.8-16.7       & max & 42868.539 &   0.496878  &         \\
 193 & 18 06 38.35 +8 02 52.2 & A2 0975-10464850 & EA     & 15.3-16.2-15.5  & min & 45232.24  &   1.51209   &         \\
 194 & 18 06 46.37 +8 50 11.6 & A2 0975-10471271 & SR     & 15.3-16.4       &     &           & 254         & 1       \\
 195 & 18 06 51.41 +5 09 30.2 & A2 0900-11608642 & EW     & 15.0-15.6-15.55 & min & 46653.414 &   0.478173  &         \\
 196 & 18 06 56.18 +6 27 48.4 & A2 0900-11614215 & RRC    & 14.3-14.8       & max & 46617.342 &   0.322833  &         \\
 197 & 18 07 05.65 +6 05 14.9 & GSC 00442-00871  & LB     & 14.4-14.9       &     &           &             & 5       \\
 198 & 18 07 12.73 +4 58 10.3 & A2 0900-11632505 & EW     & 15.9-16.45-16.4 & min & 44491.256 &   0.298816  &         \\
 199 & 18 07 16.39 +5 16 52.2 & GSC 00438-00265  & LB     & 14.7-15.15      &     &           &             & 1       \\
 200 & 18 07 21.54 +5 32 13.3 & A2 0900-11642030 & SRB    & 15.8-16.5       &     &           &  78.3:      & 1, 27   \\
 201 & 18 07 36.93 +7 26 35.9 & A2 0900-11659025 & RRAB   & 15.9-16.7       & max & 42875.531 &   0.588969  &         \\
 202 & 18 07 41.28 +6 45 28.6 & GSC 00443-00936  & SRB:   & 15.3-16.4       &     &           & 145:        & 1, 28   \\
 203 & 18 07 49.65 +5 27 51.6 & GSC 00439-03982  & EA     & 13.7-14.3-13.75 & min & 44455.302 &   0.711615  &         \\
 204 & 18 07 56.04 +4 56 46.7 & GSC 00439-01998  & EW     & 13.85-14.1-14.05& min & 44025.432 &   0.519085  &         \\
 205 & 18 07 59.01 +4 45 55.6 & GSC 00439-03124  & LB     & 14.85-15.4      &     &           &             & 1       \\
 206 & 18 08 01.92 +5 23 57.2 & GSC 00439-00910  & LB     & 14.9-15.5       &     &           &             & 1       \\
 207 & 18 08 03.10 +6 14 14.3 & A2 0900-11688498 & SRA    & 15.3-16.5       &     &           & 131         & 1, 29   \\
 208 & 18 08 10.12 +5 40 58.3 & A2 0900-11696307 & RRC:   & 15.4-16.0       & max & 46972.316 &   0.312194  &         \\
 209 & 18 08 11.17 +3 52 53.6 & A2 0900-11697505 & RRAB   & 14.3-15.2       & max & 44489.274 &   0.675244  &         \\
\hline
\end{tabular}
}

\newpage

Table 1. Continued

\medskip

{\small
\begin{tabular}{|r|c|l|l|l|cl|l|l|}
\hline MDV & Coord. (J2000) & GSC / USNO-A2.0 & type &
max-min-min II &\multicolumn{2}{|c|}{epoch JD24...} & period & rem. \\
\hline
 210 & 18 08 14.53 +4 47 59.9 & A2 0900-11701213 & EW     & 14.6-15.25-15.15& min & 42870.481 &   0.200466  & 30      \\
 211 & 18 08 24.30 +6 28 14.0 & GSC 00443-02136  & LB:    & 14.2-14.9       &     &           &             & 1, 27   \\
 212 & 18 08 24.31 +5 26 18.7 & A2 0900-11711826 & LB     & 15.4-16.2       &     &           &             & 1       \\
 213 & 18 08 30.82 +5 39 11.8 & A2 0900-11718894 & RRAB   & 15.35-16.0      & max & 43282.487 &   0.749949  &         \\
 214 & 18 08 44.55 +5 57 55.0 & GSC 00443-00420  & SR     & 15.4-15.9       &     &           & 180:        & 1       \\
 215 & 18 08 55.97 +5 12 35.7 & GSC 00439-02500  & SR:    & 14.55-15.3      &     &           &  51.7:      & 1       \\
 216 & 18 08 56.18 +5 57 10.1 & A2 0900-11747125 & EW     & 14.2-14.5-14.5  & min & 42930.401 &   0.421105  &         \\
 217 & 18 09 04.43 +7 55 37.8 & GSC 01009-02317  & RRC:   & 13.85-14.2      & max & 46973.455 &   0.436515  &         \\
 218 & 18 09 09.12 +5 04 23.6 & A2 0900-11761590 & EB     & 15.7-16.2-16.05 & min & 42892.539 &   0.97154   &         \\
 219 & 18 09 09.16 +3 47 52.1 & GSC 00439-03557  & LB     & 15.0-15.8       &     &           &             & 1       \\
 220 & 18 09 09.93 +7 28 09.8 & GSC 00443-00094  & SRB    & 14.9-16.0       &     &           &  65.7       & 1       \\
 221 & 18 09 13.54 +5 50 41.1 & A2 0900-11766734 & RRC    & 15.85-16.3      & max & 42868.539 &   0.339997  &         \\
 222 & 18 09 23.78 +6 51 47.7 & GSC 00443-00758  & LB     & 15.0-15.9       &     &           &             & 1       \\
 223 & 18 09 26.16 +5 35 58.9 & GSC 00439-00043  & EW     & 13.9-14.1-14.1  & min & 46979.464 &   0.393194  &         \\
 224 & 18 09 27.62 +4 28 50.7 & A2 0900-11782663 & HADS   & 15.4-15.9       & max & 42927.415 &   0.0610848 &         \\
 225 & 18 09 48.60 +6 00 37.6 & GSC 00443-00459  & SRB    & 15.1-15.9       &     &           &  59         & 1, 31   \\
 226 & 18 09 51.09 +5 03 47.2 & A2 0900-11808436 & EB     & 15.5-16.0-15.8  & min & 46971.317 &   0.621532  &         \\
 227 & 18 09 53.94 +4 16 08.6 & GSC 00439-00357  & EA     & 13.6-14.1       & min & 42922.49  &   1.94171   &         \\
 228 & 18 09 59.29 +4 44 29.1 & A2 0900-11817531 & RRAB   & 14.9-15.6       & max & 42872.523 &   0.607108  &         \\
 229 & 18 10 07.79 +7 58 22.7 & GSC 01009-02199  & LB     & 14.9-15.4       &     &           &             & 1       \\
 230 & 18 10 17.44 +8 11 27.1 & GSC 01009-01807  & EW     & 14.1-14.6-14.6  & min & 46979.390 &   0.449795  &         \\
 231 & 18 10 20.07 +6 02 08.0 & A2 0900-11840595 & EB     & 14.6-15.2-15.0  & min & 42927.410 &   0.654786  &         \\
 232 & 18 10 24.56 +5 27 16.3 & A2 0900-11845655 & EB     & 15.15-15.8-15.5 & min & 42870.546 &   0.396235  &         \\
 233 & 18 10 29.54 +4 22 49.3 & GSC 00439-03369  & RRC    & 14.4-14.7       & max & 43284.483 &   0.236166  & 9       \\
 234 & 18 10 55.71 +6 20 08.7 & A2 0900-11878794 & EB     & 14.7-15.45-15.1 & min & 43694.395 &   0.901138  &         \\
 235 & 18 10 59.00 +5 07 48.6 & GSC 00439-01424  & LB     & 15.1-16.0       &     &           &             & 1       \\
 236 & 18 11 05.34 +7 54 03.6 & GSC 01009-02424  & EW     & 14.0-14.3-14.3  & min & 42890.512 &   0.410104  &         \\
 237 & 18 11 12.67 +5 26 17.1 & A2 0900-11896688 & EW     & 15.0-15.5-15.4  & min & 45203.305 &   0.492401  &         \\
 238 & 18 11 20.03 +7 16 11.9 & A2 0900-11904490 & EW     & 14.9-15.25-15.2 & min & 42891.529 &   0.585531  & 3       \\
 239 & 18 11 22.94 +3 43 39.2 & GSC 00435-00252  & RRAB   & 14.8-15.4       & max & 42901.520 &   0.565079  &         \\
 240 & 18 11 25.22 +8 41 15.5 & A2 0975-10706743 & RRAB   & 14.9-15.9       & max & 42902.512 &   0.498133  &         \\
 241 & 18 11 51.49 +3 50 02.6 & A2 0900-11938410 & EW     & 15.5-16.1-16.1  & min & 46344.236 &   0.443803  &         \\
 242 & 18 12 07.63 +6 03 44.7 & GSC 00443-01265  & LB     & 14.8-15.7       &     &           &             & 1       \\
 243 & 18 12 09.75 +5 18 40.1 & GSC 00439-00431  & SR     & 15.0-15.7       &     &           &  42.4       &         \\
 244 & 18 12 14.39 +9 06 17.7 & GSC 01009-00236  & SR:    & 14.3-15.3       &     &           &             & 1, 32   \\
 245 & 18 12 21.45 +5 26 55.7 & GSC 00439-01334  & SR:    & 13.8-14.2       &     &           &             & 1, 33   \\
 246 & 18 12 29.32 +5 10 05.6 & A2 0900-11978953 & LB     & 15.6-16.4       &     &           &             & 1       \\
 247 & 18 12 31.61 +5 21 09.6 & A2 0900-11981473 & EW     & 15.7-16.25-16.2 & min & 44839.273 &   0.360409  & 34      \\
 248 & 18 12 37.32 +3 49 33.2 & A2 0900-11987723 & LB     & 15.4-16.4       &     &           &             & 1       \\
 249 & 18 12 37.92 +7 18 23.1 & A2 0900-11988370 & RRAB   & 15.1-16.2       & max & 46591.462 &   0.647362  &         \\
 250 & 18 12 40.09 +4 45 30.6 & A2 0900-11990648 & LB     & 15.0-15.6       &     &           &             & 1       \\
 251 & 18 12 59.75 +4 20 35.3 & GSC 00439-04024  & LB     & 15.2-16.3       &     &           &             & 1       \\
 252 & 18 13 00.21 +6 52 27.4 & A2 0900-12012307 & EB     & 14.8-15.4-15.2  & min & 42930.401 &   0.451938  &         \\
 253 & 18 13 01.88 +3 40 41.8 & A2 0900-12014075 & LB     & 15.7-16.3       &     &           &             & 1       \\
 254 & 18 13 06.78 +8 15 49.4 & GSC 01009-01148  & LB     & 14.7-15.5       &     &           &             & 1       \\
 255 & 18 13 08.61 +5 25 22.3 & A2 0900-12021203 & SR:    & 15.1-16.2       &     &           & 182         & 1, 35   \\
 256 & 18 13 13.90 +6 16 54.9 & GSC 00443-02710  & SRB    & 14.1-14.7       &     &           &  58:        & 1, 36   \\
 257 & 18 13 21.93 +4 20 39.6 &2M18132192+0420395& RRAB   & 14.55-15.4      & max & 44815.380 &   0.538190  &         \\
 258 & 18 13 27.91 +6 23 12.4 & GSC 00443-02477  & SR:    & 15.3-16.0       &     &           & 415:        & 1       \\
 259 & 18 13 37.28 +6 53 38.4 & GSC 00443-01862  & LB     & 14.6-15.3       &     &           &             & 1       \\
 260 & 18 13 44.99 +6 32 19.9 & A2 0900-12059839 & RRAB   & 15.8-16.3       & max & 42957.469 &   0.629529  &         \\
 261 & 18 13 54.90 +4 42 37.9 & GSC 00439-02896  & EW     & 14.6-15.4-15.3  & min & 42872.523 &   0.353219  &         \\
 262 & 18 13 56.50 +6 22 44.3 & GSC 00443-02513  & LB     & 15.5-16.0       &     &           &             & 1, 37   \\
 263 & 18 13 58.69 +8 55 44.0 & A2 0975-10845746 & EB     & 15.1-16.0-15.3  & min & 46978.31  &   1.084030  &         \\
 264 & 18 14 07.60 +9 01 41.7 & GSC 01009-00647  & SR:    & 15.1-16.3       &     &           &  81:        & 1, 38   \\
 265 & 18 14 12.34 +5 07 30.4 & GSC 00439-00952  & LB     & 14.4-15.1       &     &           &             & 1       \\
 \hline
\end{tabular}
}

\newpage

Table 1. Continued
\medskip

{\small
\begin{tabular}{|r|c|l|l|l|cl|l|l|}
\hline MDV & Coord. (J2000) & GSC / USNO-A2.0 & type &
max-min-min II &\multicolumn{2}{|c|}{epoch JD24...} & period & rem. \\
\hline
 266 & 18 14 22.84 +5 31 30.7 & A2 0900-12100102 & RRAB   & 14.9-16.5       & max & 42902.512 &   0.444778  &         \\
 267 & 18 14 22.88 +6 09 56.4 & A2 0900-12100136 & RRAB   & 15.7-16.4       & max & 44489.274 &   0.756962  &         \\
 268 & 18 14 24.38 +7 12 52.2 & GSC 00443-00593  & SRB:   & 14.9-15.8       &     &           &  69.8:      & 1       \\
 269 & 18 15 04.17 +7 55 41.5 & A2 0975-10907379 & RRAB   & 15.0-15.9       & max & 44113.304 &   0.486940  &         \\
 270 & 18 15 11.71 +6 47 02.6 & GSC 00444-00676  & RRAB   & 13.6-14.1       & max & 43199.585 &   0.614567  &         \\
 271 & 18 15 13.53 +9 06 07.8 & GSC 01009-01210  & LB     & 14.8-15.3       &     &           &             & 1       \\
 272 & 18 15 14.48 +7 29 35.4 & A2 0900-12156313 & LB     & 15.3-16.4       &     &           &             & 1       \\
 273 & 18 15 20.96 +5 39 10.4 & A2 0900-12163463 & SRB    & 15.4-16.1       &     &           &  85:        & 5       \\
 274 & 18 15 28.15 +6 47 52.9 & A2 0900-12171052 & EW     & 15.8-16.3-16.25 & min & 44397.415 &   0.363117  &         \\
 275 & 18 15 31.44 +6 19 20.1 & GSC 00444-01586  & LB     & 14.8-15.6       &     &           &             & 1       \\
 276 & 18 15 38.82 +6 29 58.9 & A2 0900-12182330 & SRB    & 14.6-15.2       &     &           &  61:        & 1       \\
 277 & 18 15 47.01 +5 38 34.6 & A2 0900-12191325 & RRC    & 15.8-16.2       & max & 46653.414 &   0.284193  & 39      \\
 278 & 18 15 47.87 +6 18 41.2 & A2 0900-12192227 & EB:    & 14.8-15.15-15.05& min & 46591.46  &   1.54528   & 40      \\
 279 & 18 15 48.94 +7 08 33.4 & A2 0900-12193398 & RRAB   & 15.5-16.4       & max & 44107.290 &   0.566656  & 41      \\
 280 & 18 16 11.22 +7 21 48.1 & A2 0900-12217660 & LB     & 15.5-16.2       &     &           &             & 1       \\
 281 & 18 16 18.43 +6 16 10.6 & GSC 00444-02004  & SR     & 14.1-15.0       &     &           &  80         & 1, 42   \\
 282 & 18 16 27.14 +6 42 55.5 & GSC 00444-00546  & SRB:   & 15.2-15.8       &     &           &  86:        &         \\
 283 & 18 16 35.19 +5 34 35.0 & A2 0900-12244463 & SR     & 15.6-16.5       &     &           & 252:        & 1       \\
 284 & 18 16 40.10 +6 37 12.3 & GSC 00444-00861  & SRB    & 15.2-16.2       &     &           &  73:        &         \\
 285 & 18 16 45.22 +7 57 50.1 & A2 0975-11002638 & RRC    & 14.8-15.5       & max & 43254.534 &   0.281676  & 9       \\
 286 & 18 16 46.19 +8 18 49.5 & GSC 01010-01418  & LB     & 13.7-14.2       &     &           &             &         \\
 287 & 18 16 50.09 +5 41 14.0 & GSC 00444-00149  & LB     & 14.9-15.5       &     &           &             & 1       \\
 288 & 18 17 00.69 +4 29 24.7 & GSC 00440-02278  & LB     & 14.4-14.9       &     &           &             & 1       \\
 289 & 18 17 11.39 +6 18 13.2 & GSC 00444-02072  & RRAB   & 14.6-15.15      & max & 42872.523 &   0.820628  &         \\
 290 & 18 17 20.57 +6 08 43.7 & A2 0900-12293340 & EW     & 15.4-16.0-15.9  & min & 46623.455 &   0.325430  &         \\
 291 & 18 17 22.37 +5 26 14.0 & GSC 00440-00741  & LB     & 15.3-16.2       &     &           &             & 1       \\
 292 & 18 17 30.45 +8 14 47.1 & A2 0975-11046328 & EA     & 14.3-14.8-14.45 & min & 43420.250 &   0.845830  &         \\
 293 & 18 17 32.09 +8 14 16.5 & A2 0975-11047913 & RRAB   & 15.2-16.2       & max & 43197.623 &   0.521403  &         \\
 294 & 18 17 37.92 +4 58 12.4 & A2 0900-12311653 & EW     & 15.1-15.7-15.6  & min & 44131.297 &   0.423964  &         \\
 295 & 18 18 00.35 +5 18 06.2 & A2 0900-12334379 & EA     & 15.3-16.3:-16.0:& min & 43198.60  &   3.05819   & 43      \\
 296 & 18 18 31.41 +4 15 21.9 & GSC 00440-01122  & SR:    & 15.1-16.1       &     &           &             & 1, 44   \\
 297 & 18 18 56.33 +4 40 05.5 & GSC 00440-01831  & SR     & 14.4-16.0       &     &           & 148         & 1       \\
 298 & 18 18 57.01 +6 37 53.5 & GSC 00444-01364  & LB     & 14.35-14.8      &     &           &             & 5       \\
 299 & 18 19 17.22 +4 57 27.5 & A2 0900-12418606 & RRAB   & 15.1-16.0       & max & 43272.409 &   0.534392  &         \\
 300 & 18 19 18.43 +6 34 41.3 & GSC 00444-01143  & EW     & 15.0-15.3-15.3  & min & 43422.199 &   0.457672  &         \\
 301 & 18 19 20.32 +4 39 48.0 & A2 0900-12422241 & EW     & 15.1-15.8-15.7  & min & 44732.521 &   0.376235  &         \\
 302 & 18 19 21.17 +5 17 20.1 & GSC 00440-00611  & SR:    & 15.1-15.9       &     &           &             & 1, 28   \\
 303 & 18 19 21.38 +6 22 45.3 & A2 0900-12423516 & EB     & 15.35-15.9-15.6 & min & 43277.523 &   0.938567  &         \\
 304 & 18 19 28.92 +5 37 54.0 & A2 0900-12432214 & RRAB   & 15.4-16.3       & max & 46973.455 &   0.762271  &         \\
 305 & 18 19 34.90 +6 12 10.2 & A2 0900-12438878 & SR:    & 15.15-15.6      &     &           &  88.3:      &         \\
 306 & 18 19 52.67 +4 16 36.6 & GSC 00440-02850  & LB     & 14.3-14.9       &     &           &             & 1       \\
 307 & 18 20 00.86 +8 25 09.5 & A2 0975-11206579 & EW:    & 15.3-15.9-15.9  & min & 44839.273 &   0.527540  & 3       \\
 308 & 18 20 04.10 +4 11 34.4 & A2 0900-12469784 & EW     & 14.9-15.5-15.4  & min & 43422.199 &   0.465738  &         \\
 309 & 18 20 15.58 +8 03 36.4 & GSC 01010-02424  & LB     & 14.5-15.3       &     &           &             & 1       \\
 310 & 18 20 19.35 +6 20 05.9 & A2 0900-12484774 & LB     & 15.5-16.1       &     &           &             & 1       \\
 311 & 18 20 30.08 +3 48 02.5 & A2 0900-12494927 & HADS   & 15.2-15.8       & max & 43243.438 &   0.097296  & 45      \\
 312 & 18 20 55.11 +4 44 46.0 & A2 0900-12518388 & EW     & 15.1-15.7-15.65 & min & 46646.401 &   0.463428  &         \\
\hline
\end{tabular}
}
\newpage

Table 2. New Data on Known Variables
\medskip

{\small

\begin{tabular}{|l|l|c|l|l|cl|l|l|}
\hline GCVS/NSV & HV/SON & Coord. (J2000) & type & max-min-min II
&\multicolumn{2}{|c|}{epoch JD24...} & period & rem. \\ \hline
NSV 9475  & HV 11011 & 17 40 13.31 +6 02 51.8 & RRAB & 15.4-16.4       & max & 42957.370 &   0.526203 &    \\
NSV 9642  & HV 11040 & 17 45 26.26 +8 22 01.8 & RRAB & 14.6-16.0       & max & 42870.481 &   0.467784 &    \\
NSV 9704  & HV 11046 & 17 48 06.28 +8 12 54.2 & RRC  & 14.1-14.7       & max & 42934.380 &   0.320757 &    \\
NSV 9721  & S 9837   & 17 49 03.37 +5 06 19.3 & EW   & 15.4-16.0-15.9  & min & 42930.509 &   0.255458 &    \\
NSV 9734  & HV 11053 & 17 49 30.19 +4 18 40.1 & LB   & 14.9-15.6       &     &           &            & 1  \\
NSV 9740  & S 9838   & 17 49 43.48 +4 13 24.1 & EA   & 15.1-15.9-15.2: & min & 44112.30  &   1.86895  &    \\
NSV 9973  & S 9277   & 18 00 32.19 +5 27 11.3 & LB   & 15.2-16.0       &     &           &            & 1  \\
V947 Oph  & S 4199   & 18 02 05.31 +5 52 45.5 & EA   & 14.3-14.8       & min & 44023.455 &   0.797747 &    \\
NSV 10129 & S 9857   & 18 03 54.74 +7 34 27.4 & SR:  & 14.7-15.4       &     &           & 143:       & 1  \\
NSV 10291 & S 9867   & 18 09 52.67 +3 41 59.3 & LB   & 15.2-15.9       &     &           &            & 1  \\
V2087 Oph & S 9297   & 18 11 16.36 +5 15 32.3 & RRAB & 15.1-16.3       & max & 43282.452 &   0.495589 &    \\
NSV 10381 & S 9298   & 18 13 09.74 +4 28 58.1 & LB   & 14.0-14.7       &     &           &            & 1  \\
NSV 10403 & S 9872   & 18 14 00.45 +3 50 35.0 & LB   & 14.7-15.1       &     &           &            & 1  \\
\hline \end{tabular}

}

\bigskip
\bigskip
\end{center}

\hoffset=0.0\textwidth

Remarks to the Tables 1 and 2.

1. Variable in NSVS data. 2. $P = 55.6$ d (from NSVS
data). 3. A twice shorter period and type RRC are possible. 4. 
$P = 39.7$ d (NSVS data). 5. A small-amplitude variable in
NSVS data. 6. $P = 60$ d (NSVS data). 7. $P\sim 50$ d
(NSVS data) is possible. 8. A twice longer period is possible. 9.
A twice longer period and type EW are possible. 10. $P\sim 51$ d
(NSVS data) is possible. 11. A CCD study following this
discovery was announced in Antipin et al. (2007). 12. A white or
yellow star, $J - K = 0.529$ (2MASS). 13. $P \sim 62$ d
(NSVS data) is possible. 14. NSVS data show variations with the
same period. 15. 1-day aliases of a twice shorter period are
strong. 16. $P = 45$ d (NSVS data). 17. Variable in ASAS-3
data, not included into the ASAS-3 catalog of variable stars. 18.
Not identical to V568 Oph ($17^{\rm h}59^{\rm m}44.\!\!^{\rm
s}09$, $+4^\circ59'55.\!\!^{\prime\prime}6$, J2000). 19. 1-day
aliases (0.399278 d and 0.285235 d) are
also quite possible. 20. The periods 83.8 d or
92.5 d are possible (NSVS data). 21. A double star.
22. $P \sim 63.5$ d (NSVS data) is possible. 23. $P =
75$ d (NSVS data). 24. The period 0.286864 d
(type EW) is also quite possible. 25. O'Connell effect. 26. A
1-day alias, $P = 0.49813$ d, is possible. 27. $P\sim
78$ d (NSVS data) is possible. 28. $P \sim 150$ d
(NSVS data) is possible. 29. $P \sim 130$ d (NSVS data) is
possible. 30. A twice shorter period and type HADS are possible.
31. $P = 58$ d (NSVS data). 32. $P = 48$ d (NSVS
data). 33. $P = 62$ d (NSVS data). 34. $P = 0.305252$ d
is also possible. 35. $P \sim 250$ d (NSVS data)
is possible. 36. $P \sim 60$ d (NSVS data) is possible. 37.
$P \sim 54$ d (NSVS data) is possible. 38. $P \sim 85$ d
(NSVS data) is possible. 39. A 1-day alias, $P = 0.397451$ d,
is possible. 40. A twice shorter period and type RRAB
are possible. 41. The coordinates are from the USNO-A2.0
catalogue. 42. $P = 81$ d (NSVS data). 43. Possibly,  the
minima are deeper. 44. $P \sim 82$ d (NSVS data) is
possible. 45. A 1-day alias, $P = 0.081429$ d, is
possible.

\bigskip
\bigskip

\begin{center}

REFERENCES

\end{center}

\medskip

Antipin, S.V., {\it et al.} 2007, {\it Perem. Zvezdy}, {\bf 27},
No.~8

Antipin, S.V., Shugarov, S.Yu., and Kroll, P. 2002, {\it Inform.
Bull. Var. Stars}, No.~5246

Bertin, E., and Arnouts, S. 1996, {\it Astron. Astrophys., Suppl.
Ser.}, {\bf 117}, 393

Brunzendorf, J., and Meusinger, H. 2001, {\it Astron. Astrophys.},
{\bf 373}, 38

Cutri, R.M., {\it et al.} 2003, {\it The 2MASS All-Sky Catalog of
Point Sources} (CDS, II/246)

Devor, J. 2005, {\it Astrophys. J.}, {\bf 628}, 411

G\"otz, W., Huth, H., and Hoffmeister, C. 1957, {\it Ver\"off.
Sternw. Sonneberg}, {\bf 4}, H.~2

Henze, M., Meusinger, H., and Pietsch, W. 2008, {\it Astron.
Astrophys.}, {\bf 477}, 67

Kolesnikova, D., {\it et al.} 2007a, {\it Perem. Zvezdy Suppl.},
{\bf 7}, No.~3

Kolesnikova, D., {\it et al.} 2007b, {\it Perem. Zvezdy Suppl.},
{\bf 7}, No.~24

Kov\'acs, G., Zucker, S., and Mazeh, T. 2002, {\it Astron.
Astrophys.}, {\bf 391}, 369

Manannikov, A.L., {\it et al.} 2006, {\it Perem. Zvezdy Suppl.},
{\bf 6}, No.~34

Monet, D., {\it et al.}, 1998, {\it USNO-A V2.0. A Catalog of
Astrometric Standards} (Washington: USNO)

Pojmanski, G. 2002, {\it Acta Astron.}, {\bf 52}, 397

Samus, N.N. 1983, {\it Mitt. Ver\"and. Sterne}, {\bf 9}, 87

Samus, N.N., {\it et al.} 2006, in {\it ``Virtual Observatory:
Plate Content Digitization, Archive Mining, Image Sequence
Processing'', Proc. Internat. Workshop}, M.~Tsvetkov et al.
(eds.), Sofia: Heron Press, p. 103

Scholz, R.-D., Meusinger, H., and Irwin, M. 1997, {\it Astron.
Astrophys.}, {\bf 325}, 457

Schwarzenberg-Czerny, A. 1989, {\it Astron. Astrophys.}, {\bf
210}, 174

Schwarzenberg-Czerny, A. 1996, {\it Astroph. J.}, {\bf 460}, L107

Sokolovsky, K.V. 2006, {\it Perem. Zvezdy Suppl.}, {\bf 6}, No.~18

Sokolovsky, K., and Lebedev, A. 2005, in {\it ``12th Young
Scientists' Conference on Astronomy and Space Physics, Kyiv,
Ukraine, April 19-23, 2005''}, A.~Simon and A.~Golovin (eds.),
p.~79

Vogt, N., Kroll, P., and Splittgerber, E. 2004, {\it Astron.
Astrophys.}, {\bf 428}, 925

Welch, D.L., and Stetson, P.B. 1993, {\it Astron. J.}, {\bf 105},
1813

Wo\'zniak, P.R., {\it et al.} 2004, {\it Astron. J.}, {\bf 127},
2436

\newpage
\bigskip
\Huge
\begin{center}

{\bf Online material}

\end{center}

\begin{figure}
\center{\includegraphics[angle=0,width=1.0\textwidth]{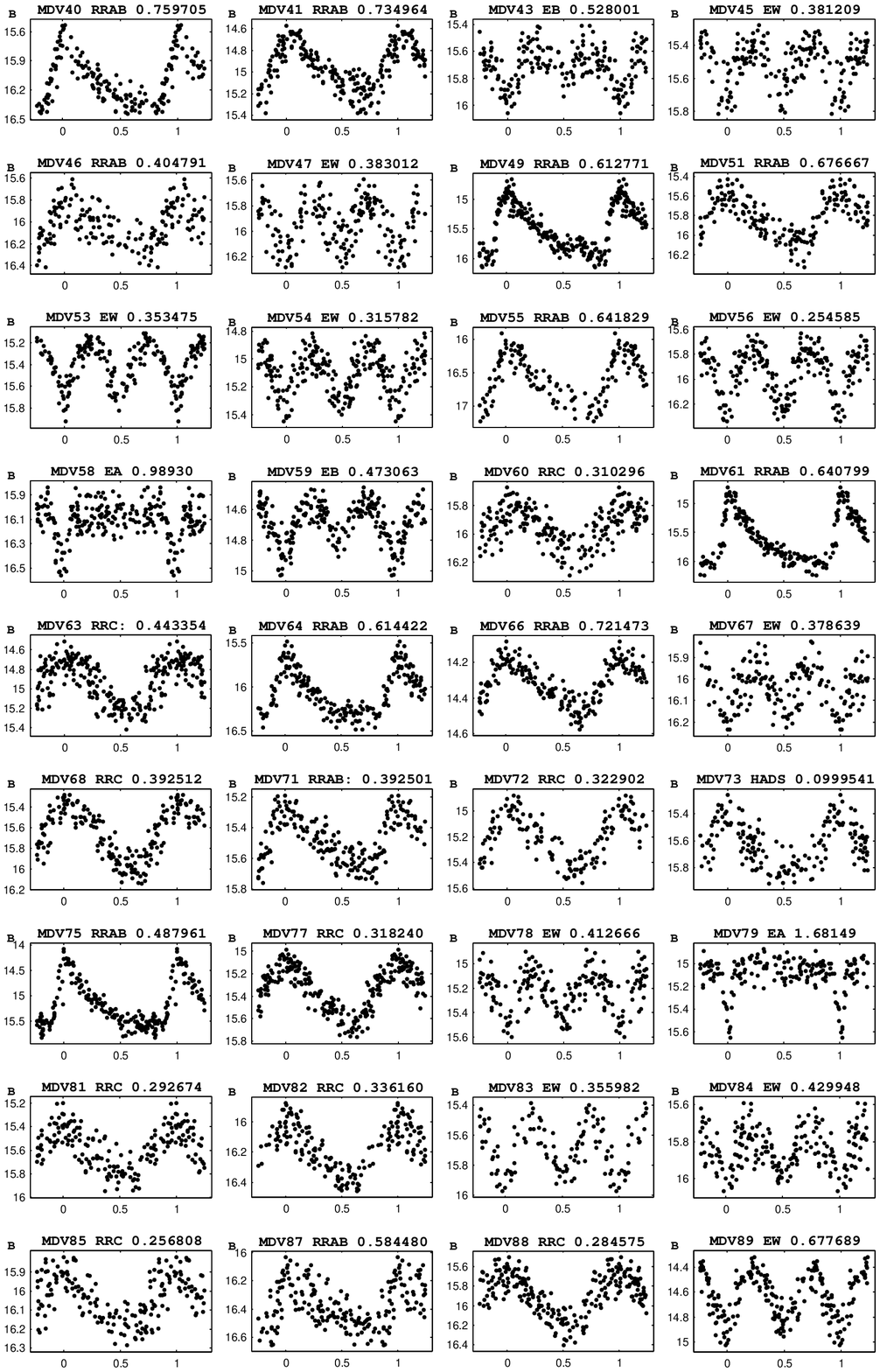}}
\caption{Phased light curves of the new regular
variable stars.}
\end{figure}

\setcounter{figure}{5}
\begin{figure}
\center{\includegraphics[angle=0,width=1.0\textwidth]{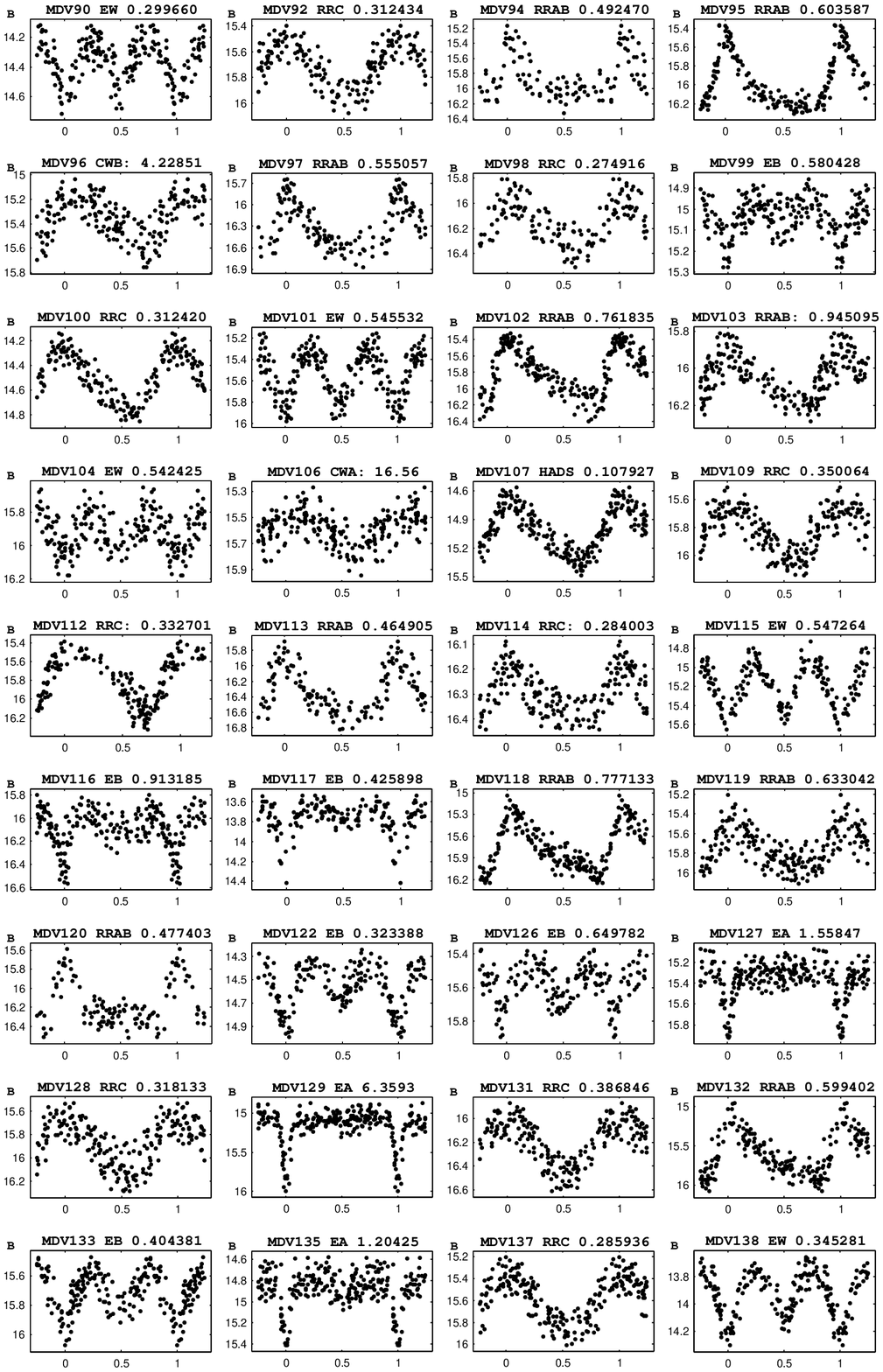}}
\caption{Continued}
\end{figure}

\setcounter{figure}{5}
\begin{figure}
\center{\includegraphics[angle=0,width=1.0\textwidth]{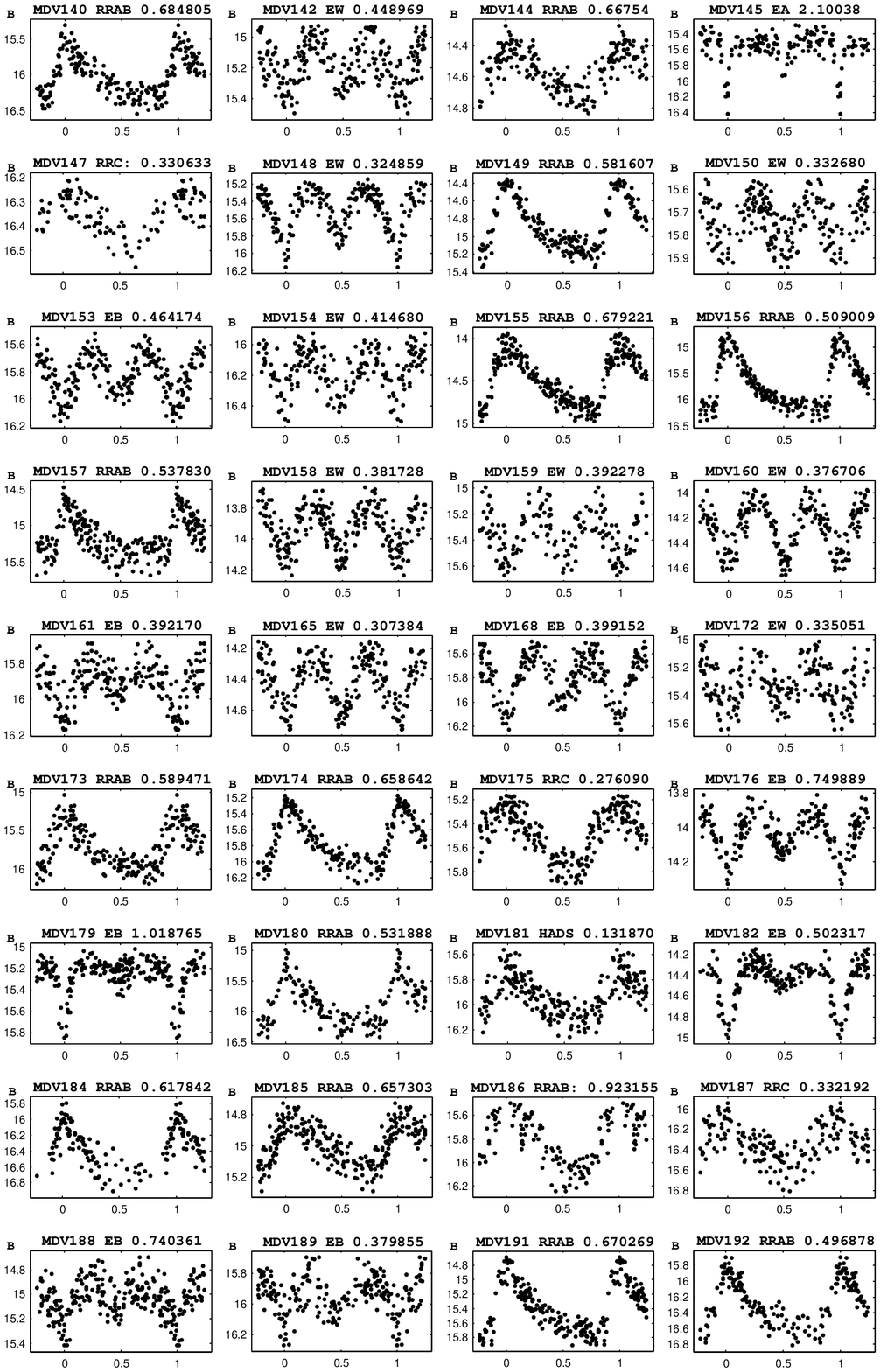}}
\caption{Continued}
\end{figure}

\setcounter{figure}{5}
\begin{figure}
\center{\includegraphics[angle=0,width=1.0\textwidth]{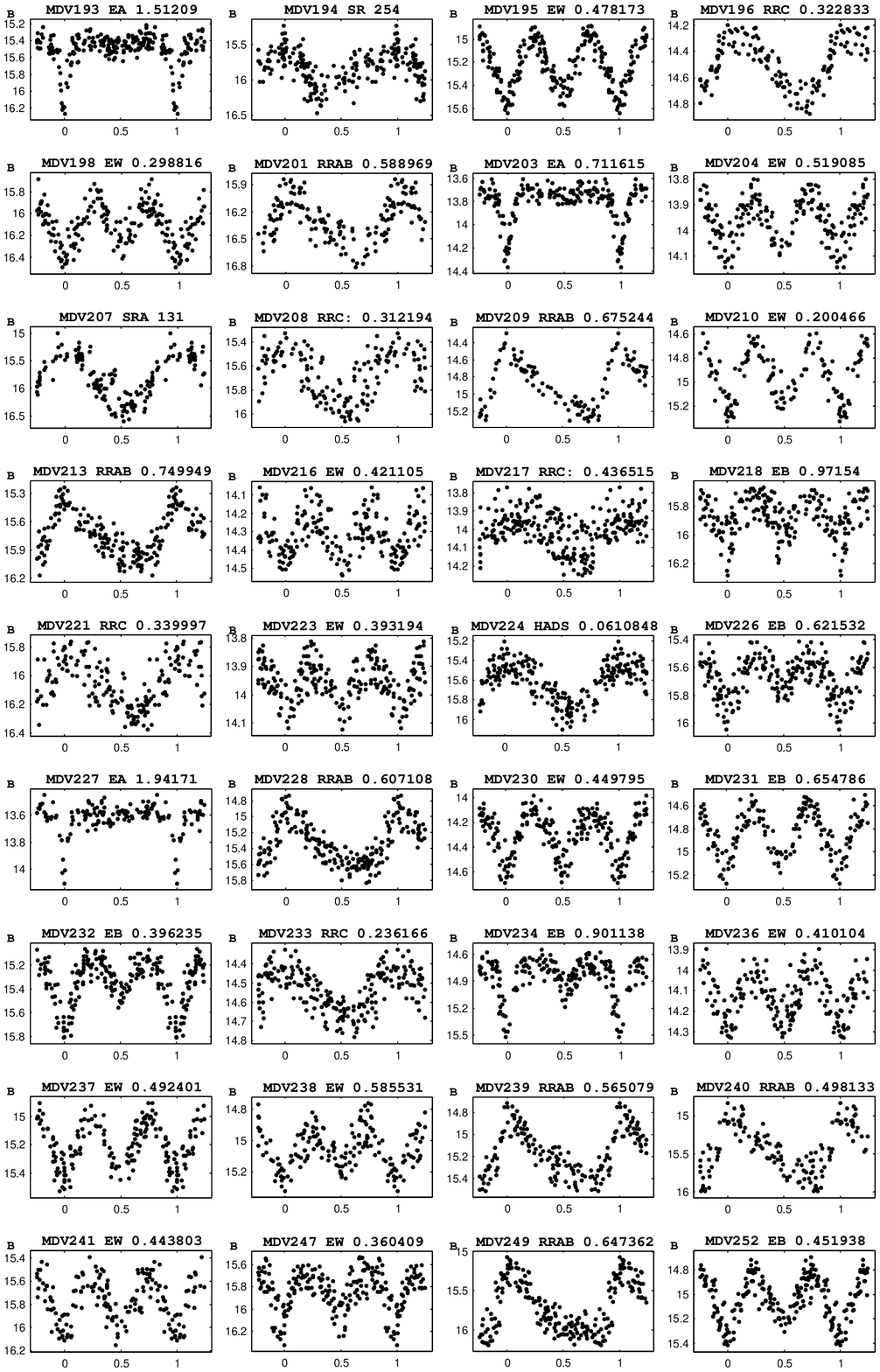}}
\caption{Continued}
\end{figure}

\setcounter{figure}{5}
\begin{figure}
\center{\includegraphics[angle=0,width=1.0\textwidth]{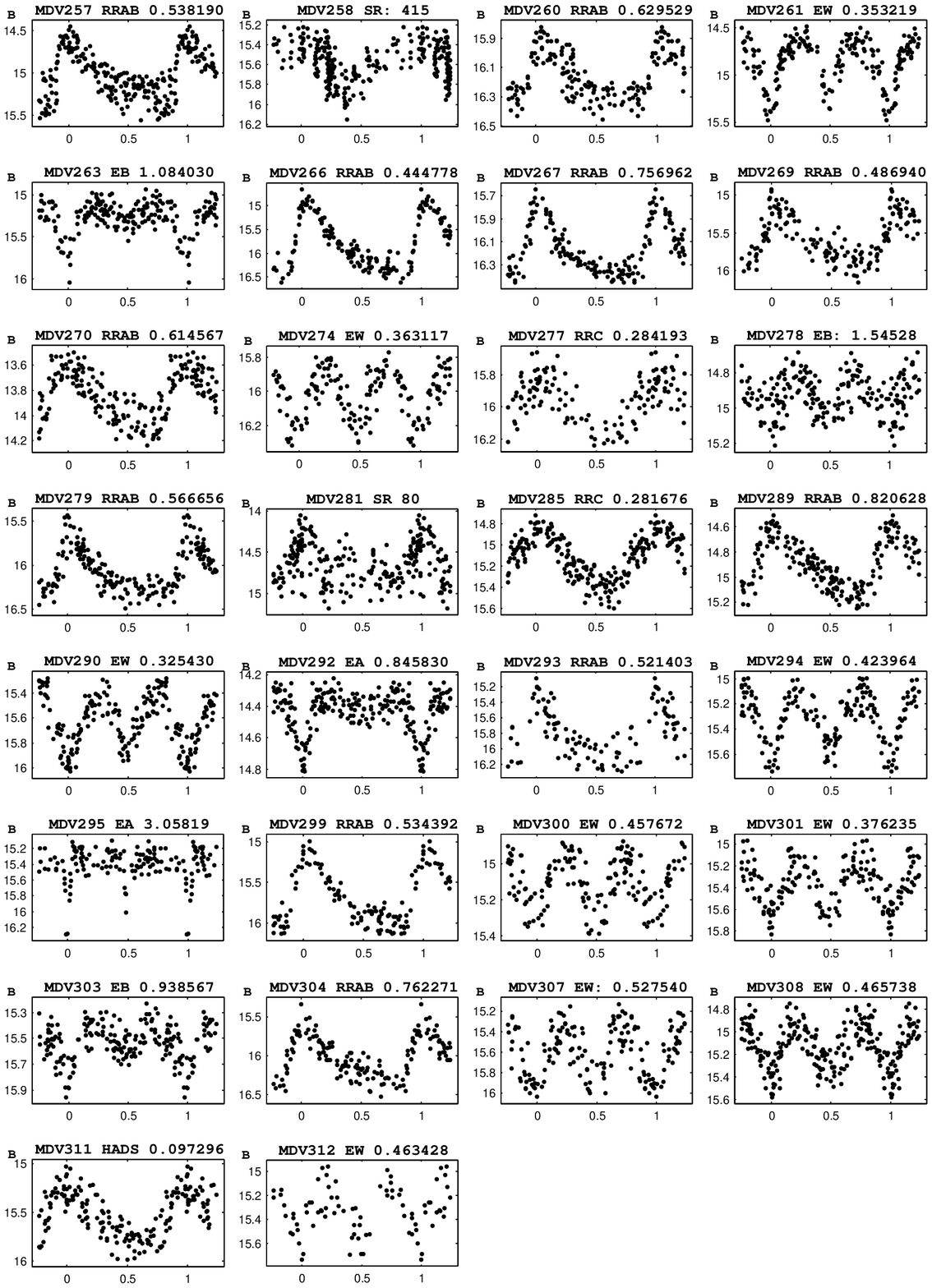}}
\caption{Continued}
\end{figure}

\end{document}